\documentclass[journal,comsoc,twocolumn]{IEEEtran}
\usepackage[utf8]{inputenc}
\usepackage[T1]{fontenc}
\usepackage{float}
\usepackage{amsmath}
\usepackage{amsfonts}
\usepackage{mathtools}
\usepackage{amsfonts}
\IEEEoverridecommandlockouts
\usepackage{amssymb}
\usepackage{graphicx}
\usepackage{epstopdf}
\usepackage{makeidx}
\usepackage{csquotes}
\usepackage{graphicx}
\usepackage{cite}
\usepackage{xcolor}
\usepackage{subcaption}
\usepackage{graphicx}
\usepackage{balance}

\begin{document}

\title{Transmissive Beyond Diagonal RIS-Mounted LEO Communication for NOMA IoT Networks}
%NOMA-Enabled Transmissive Beyond Diagonal RIS Mounted LEO Satellite Communication
\author{Wali Ullah Khan, Eva Lagunas, Symeon Chatzinotas \\Interdisciplinary Centre for Security, Reliability and Trust (SnT), University of Luxembourg\\
\{waliullah.khan, eva.lagunas, symeon.chatzinotas\}@uni.lu
%\thanks{This work has been supported by the Luxembourg National Research Fund (FNR) under the project MegaLEO (C20/IS/14767486).}
}%

\markboth{IEEE Wireless Communications and Networking Conference, 2025
}
{Shell \MakeLowercase{\textit{et al.}}: Bare Demo of IEEEtran.cls for IEEE Journals} 

\maketitle

\begin{abstract}
Reconfigurable Intelligent Surface (RIS) technology has emerged as a transformative solution for enhancing satellite networks in next-generation wireless communication. The integration of RIS in satellite networks addresses critical challenges such as limited spectrum resources and high path loss, making it an ideal candidate for next-generation Internet of Things (IoT) networks. This paper provides a new framework based on transmissive beyond diagonal RIS (T-BD-RIS) mounted low earth orbit (LEO) satellite networks with non-orthogonal multiple access (NOMA). The NOMA power allocation at LEO and phase shift design at T-BD-RIS are optimized to maximize the system's spectral efficiency. The optimization problem is formulated as non-convex, which is first transformed using successive convex approximation and then divided into two problems. A closed-form solution is obtained for LEO satellite transmit power using KKT conditions, and a semi-definite relaxation approach is adopted for the T-BD-RIS phase shift design. Numerical results are obtained based on Monte Carlo simulations, which demonstrate the advantages of T-BD-RIS in satellite networks.
\end{abstract}


% For peerreview papers, this IEEEtran command inserts a page break, and
\IEEEpeerreviewmaketitle

\section{Introduction}
Satellite communication has become a cornerstone technology for providing reliable and wide-ranging connectivity in an increasingly interconnected world \cite{10500741}. Unlike terrestrial communication setups, satellite communication is inherently ubiquitous, offering global coverage that transcends geographical barriers such as mountains, oceans, and remote terrains \cite{10399870}. This capability enables seamless connectivity in underserved and hard-to-reach regions, including rural areas, polar zones, and maritime or aeronautical environments. With their wide area coverage and independence from terrain, satellites play a critical role in ensuring communication continuity where terrestrial infrastructure is impractical or economically unfeasible \cite{10574260}.

The advantages of satellite communication extend beyond coverage. Their broadcast capability supports efficient point-to-multipoint communication, enabling applications such as television, radio, and emergency alerts \cite{10494748}. Additionally, the resilience of satellite communication to natural disasters and their ability to provide connectivity on the move make them indispensable for disaster recovery, mobile platforms, and global navigation systems \cite{10600459}. As advancements in satellite technologies reduce latency and improve spectral efficiency, satellite communication is poised to address emerging demands in remote sensing, the Internet of Things, and 6G networks, solidifying their position as key enablers of modern communication ecosystems \cite{10489835}.

Satellite communication in low earth orbit (LEO) offers significant benefits in various areas. Due to their closeness to the ground, LEO satellite communication provides minimal delay compared to upper layer satellites (MEO and GEO), making it well-suited for time-sensitive tasks such as voice calls, video conferencing, and gaming \cite{10097680}. As the IoT devices (IoTDs) grow across smart cities, LEO communication enables real-time data collecting, analysis, and decision-making. Due to their fast response times and high data transfer speeds, LEO networks are reliable for smart city services, including traffic management, smart grid control, air quality monitoring, and emergency response \cite{10492466}. Besides the great potential, LEO communication also has some shortcomings. For example, LEO communication suffers from a weak link budget due to tiny aperture antennas at ground Internet of Thing devices (IoTDs), especially at higher frequencies like Ku-band and Ka-band \cite{9698051}. Due to their tiny size and portability, mobile device antennas inherently lose gain, reducing their signal reception and transmission. Smaller antennas and higher frequencies affect the connection budget, which compensates signal amplification and attenuation. At these frequencies, atmospheric phenomena such as water vapor, oxygen absorption, and rain interfere more with communications \cite{10286242}. This increases transmission line signal loss and reduces communication reliability. 

To address the difficulties caused by tiny aperture antennas and attenuation due to other factors in satellite communication, the implementation of Reconfigurable Intelligent Surfaces (RIS) offers a favorable option \cite{9779261}. By carefully placing RIS on tall structures like high building walls, rooftops, and other elevated sites or mounting it over the satellite, these surfaces can help efficiently guide and improve the transmission of signals from LEO to IoTDs. RIS technology facilitates the manipulation and regulation of electromagnetic waves, enabling adaptive beamforming and reduction of multipath interference \cite{10396846}. Through the strategic manipulation of the phase and amplitude of reflected signals, RIS has the capability to amplify signal strength, counteract the loss of signal along a path, and reduce the impact of atmospheric weakening \cite{9520318}. Moreover, the use of RIS in urban settings can utilize the existing infrastructure to establish virtual arrays, significantly expanding the range and coverage of LEO satellite networks. In general, RIS can be classified into diagonal RIS and beyond diagonal RIS (BD-RIS), based on the structure of the phase shift matrices they provide. In each classification, it can operate in reflective mode, transmissive mode, and hybrid mode \cite{10817282,10584518}.

It has been proven that BD-RIS performs better than conventional diagonal RIS in terms of beamforming and system capacity but at a cost of circuitry complexity \cite{10716670, 10715713}. BD-RIS offers the capability to function as a transmitter without requiring complex signal processing, distinguishing it from traditional multi-antenna systems that rely on sophisticated RF modules and entail high hardware costs \cite{10133841,khan2024integration}. Moreover, among the different operating modes of BD-RIS, transmissive BD-RIS (T-BD-RIS) holds several advantages over reflective BD-RIS. One of the main issues in reflective BD-RIS is self-interference due to the coexistence of the feed antenna and receiver on the same side, resulting in signal overlap, whereas T-BD-RIS positions the feed antenna and receivers on opposite sides, eliminating such interference \cite{9133266}. Besides that, T-BD-RIS supports wider operational bandwidth and improves aperture efficiency compared to its reflective BD-RIS \cite{9983541}. Motivated by this, integrating T-BD-RIS into satellite communication systems has the potential to significantly enhance capacity, energy efficiency, quality of service, and spectrum utilization, making it a promising solution for next-generation wireless communication systems.

Due to the potential of RIS technology in satellite communication, it has become a hot topic, and several works have carried out in the literature. In \cite{9539541}, the authors have used RIS assisted satellite networks to overcome the issue of path loss due to long transmission distances. The work in \cite{10365519} has optimized the energy efficiency of RIS-enhanced satellite networks by joint optimization of transmit power and passive beamforming. Another work in \cite{10530613} has maximized the energy efficiency of RIS enhanced NOMA satellite networks by optimizing the transmit power of the satellite and phase shift of RIS using alternating optimization. Niu {\em et al.} \cite{9897065} have proposed secure satellite communication by optimizing the beamforming and artificial noise at the transmitter and phase shift design at RIS. Wu {\em et al.} \cite{10445520} have enhanced the ergodic capacity of satellite systems by optimizing the unmanned aerial vehicle trajectory, transmit beamforming, and phase shift design. Furthermore, Pala {\em et al.} \cite{10768935} have proposed RIS assisted integrated sensing and communication in satellite networks. Of late, Asif {\em et al.} \cite{10689360} have proposed a transmissive RIS mounted LEO satellite communication to maximize the sum rate by optimizing the transmit power and passive beamforming. Moreover, Khan {\em et al.} \cite{khan2024cr}  have proposed a cognitive radio approach in two-tier satellite networks using NOMA and transmissive RIS. They maximized the sum rate of the secondary LEO network by optimizing the transmit power and phase shift design under the interference temperature threshold of the primary satellite network.

Although several works have been done on RIS enhanced LEO satellite communication, open research gaps still exist that need to be investigated. For instance, most of the existing research works focus on RIS in LEO communication with reflective RIS using diagonal phase shift matrices, and they do not consider BD-RIS. A few works consider T-RIS mounted LEO communication, but they do not study T-BD-RIS. To the best of our knowledge, the topic of NOMA LEO satellite communication using T-BD-RIS has not been studied yet. To bridge this open gap, we propose a new framework based on T-BD-RIS mounted LEO satellite IoT network with downlink NOMA. In particular, the proposed framework simultaneously optimizes the transmit power of LEO and the phase shift design of T-BD-RIS to maximize the spectral efficiency of the system. The problem is formulated as non-convex and transformed using the successive convex approximation (SCA) method. Then, it is divided into two problems for the transmit power of LEO and the phase shift design of T-BD-RIS. A closed-form solution is achieved for power allocation through KKT conditions, and a semi-definite relaxation (SDR) is adopted for phase shift design of T-BD-RIS. The rest of this work is organized as follows. In the second section, we study system and channel models. In the third section, we formulate an optimization problem and explain the proposed solution. In the fourth section, we provide numerical results and their discussion. Finally, we conclude this work.
%%%%%%%%%%%%%%%%
\begin{figure}[!t]
\centering
\includegraphics [width=0.45\textwidth]{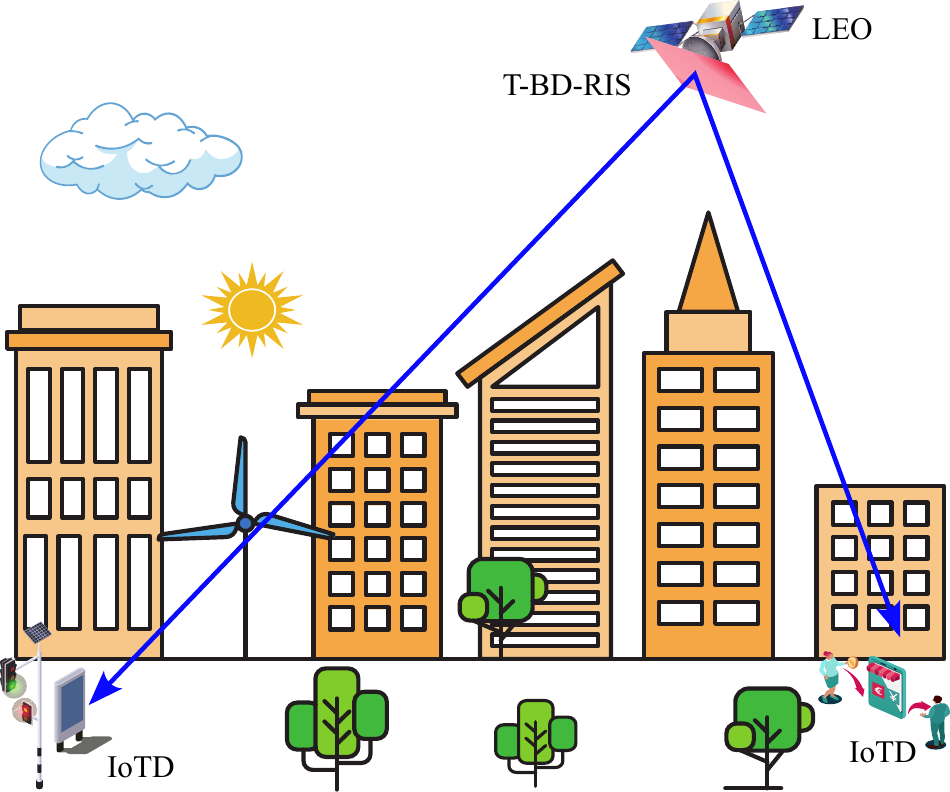}
\caption{System model}
\label{SM}
\end{figure}
%%%%%%%%%%%%%
\section{System Model}
This section explains the proposed system and channel models. We consider a downlink communication, where a $K$ elements T-BD-RIS mounted LEO satellite transmits signals to multiple IoTDs using downlink NOMA protocol, as shown in Fig. \ref{SM}. We consider a single carrier communication in this work such that the satellite accommodates only two NOMA IoTDs at any given time. For simplicity, we denote the two IoTDs as $U_i$ and $U_j$.
Let $\boldsymbol{\Phi}_t\in\mathcal C^{K_x\times K_y}$ be the phase shift matrix of T-BD-RIS such that $\boldsymbol{\Phi}_t\boldsymbol{\Phi}_t^H={\bf I_\eta}$, $\eta\in\{i,j\}$, where $K_x$ is the number of elements in each column and $K_y$ is the elements in each row, respectively. This work assumes that the channel state information of $U_i$ and $U_j$ is available at LEO. The transmit superimposed signal of LEO for $U_i$ and $U_j$ can be written as $x=\sqrt{p_iP_t}x_i+\sqrt{p_jP_t}x_j$, where $p_i$ and $p_j$ are the allocated power coefficients of $U_i$ and $U_j$ while $x_i$ and $x_j$ are their unit power signals and $P_t$ is the total transmit power of LEO satellite. It is important to mention that $p_i+p_j\leq 1$. The received signal of $U_i$ and $U_j$ can be then expressed as:
\begin{align}
y_i={\bf h}_i\boldsymbol{\Phi}_tx+n_i,\\
y_j={\bf h}_j\boldsymbol{\Phi}_tx+n_j,
\end{align}
where ${\bf h}_i\in\mathcal C^{K\times 1}$ and ${\bf h}_j\in\mathcal C^{K\times 1}$ are the channel vectors from T-BD-RIS mounted LEO to $U_i$ and $U_j$. Furthermore, $n_i$ and $n_j$ are the additive white Gaussian noise (AWGN) of $U_i$ and $U_j$. For the channel vectors from the LEO satellite to the IoTDs, we adopt a block-faded channel model expressed as:
\begin{equation}
    {\bf h}_\eta = \sqrt{\frac{g_\eta}{d_\eta^\alpha}} \hat{\bf h}_\eta e^{\vartheta\pi\psi},
\end{equation}
where \( g_\eta \) is the coefficient of small-scale Rayleigh fading, \( d_\eta \) is the distance between the satellite and user \( U_\eta \), \( \alpha \) is the path loss exponent, \( \hat{\bf h}_\eta \) is the vector of complex-valued channel gains, \( \vartheta = \sqrt{-1} \) denotes the imaginary unit, and \( \psi \) represents the Doppler shift caused by the relative motion between the satellite and user. Considering the large and small-scale fading, $\hat{\bf h}_\eta$ can be further expressed as:  
\begin{align}
&\hat{\bf h}_\eta=[1,e^{-j\rho \sin{\theta}_\eta\cos{\varphi}_\eta},\dots, e^{-j\rho \sin{\theta}_\eta\cos{\varphi}_\eta(K_x-1)}]^T\nonumber\\
&\otimes [1,e^{-j\rho \sin{\theta}_\eta\cos{\varphi}_\eta},\dots, e^{-j\rho \sin{\theta}_\eta\cos{\varphi}_\eta(K_y-1)}]^T,
\end{align}
where $\rho=2\pi f_c d_0/c$ such that $c$ is the speed of light, $f_c$ is the carrier frequency, and $d_0$ is the spacing between elements on the T-BD-RIS. Moreover, $\theta$ is the vertical and $\varphi$ is the horizontal angle of departure to $U_\eta$. For efficient implementation of SIC at receiver side, we assume that the channel gain of $U_i$ is stronger than $U_j$. Therefore, $U_i$ applies SIC to subtract the signal of $U_j$ before decoding its desired signal. However, $U_j$ cannot apply SIC and decode the signal by treating the signal of $U_i$ as a noise. Based on these observation, the rate of $U_i$ and $U_j$ can be expressed as $R_i=\log_2(1+\gamma_i)$ and $R_j=\log_2(1+\gamma_j)$. Please note that $\gamma_i$ and $\gamma_j$ are the signal to interference plus noise ratios which can be described as:
\begin{align}
\gamma_i=\frac{|{\bf h}_i\boldsymbol{\Phi}_t|^2p_iP_t}{\sigma^2},\label{5}\\
\gamma_{j}=\frac{|{\bf h}_j\boldsymbol{\Phi}_t|^2p_jP_t}{\sigma^2+|{\bf h}_j\boldsymbol{\Phi}_t|^2p_iP_t},\label{6}
\end{align}
where $\sigma^2$ in (\ref{5}) and (\ref{6}) is the variance of AWGN while the second term in the denominator of (\ref{6}) is the NOMA interference from the signal of $U_i$. 

\section{Problem Formulation and Proposed Solution}
This section provides the mathematical problem formulation and solution of the proposed system model in previous section.
\subsection{Problem Formulation}
This work aims to enhance the performance gain of NOMA LEO satellite communication using BD-RIS, which is evaluated in terms of system spectral efficiency. In particular, the proposed framework maximizes the achievable spectral efficiency of LEO communication by simultaneously optimizing the power allocation of LEO according to the downlink NOMA protocol and phase shift design of BD-RIS while ensuring the QoS of IoTDs. The joint problem of transmitting power and phase shift to maximize the spectral efficiency of the system can be formulated as follows:
%%%%%%%%%%%%%%%%%%%%%%%%%%%\\
%${\bf R}^{-1}_{\Bar{k}}{\bf R}_{k}=(1+SNR)$
\begin{alignat}{2}
\mathcal P_0:\begin{cases}  \underset{{\bf p},\boldsymbol{\Phi}_t}{\text{max}}\ (R_i+R_j)\label{7}\\
\text{subject to:} \\
C_1:\ R_{\eta}\geq R_{min},\ \eta\in i,j,\\
C_2:\ \boldsymbol{\Phi}_t\boldsymbol{\Phi}_t^H={\bf I}_K,\\
C_3:\ p_i+p_j\leq 1,\\
C_4:\ P_t(p_i+p_j)\leq P_{max},
\end{cases}
\end{alignat}
%%%%%%%%%%%%%%%%%%%%%%%%%%
where ${\bf p}\in p_i,p_j$. Constraint $C_1$ ensures the QoS of IoTDs; constraint $C_2$ invokes BD-RIS phase shift design; and constraints $C_3$ and $C_4$ control the LEO transmit power according to the NOMA principle. Moreover, $R_{min}$ is used as the thresholds of minimum rate required for QoS.
 
The optimization problem $\mathcal P_0$ is non-convex due to the rate expressions in the objective function and constraint $C_1$. Moreover, the problem is also coupled over two variables, i.e., power allocation and phase shift design. To reduce the complexity and make the optimization more tractable, we obtain a local optimal yet efficient solution, which is carried out in three steps. In the first step, we adopt successive convex approximation (SCA), which reduces the complexity of the objective function and constraint $C_1$. In the second step, we decouple the problem into separate problems for the power allocation of the LEO satellite and the phase shift of BD-RIS. In the third step, for a given phase shift design of BD-RIS, we first calculate the closed-form solution of LEO NOMA power allocation. Then, given the optimal power allocation, we design a phase shift of BD-RIS. By applying SCA on to $P_0$, the rate expression of $U_\eta$ in the objective function and $C_1$ can be re-expressed as:
%\begin{align}
%\Bar{R}_\eta= \varPi_\eta\log_2(\gamma_\eta)+(\log_2(1+\gamma_\eta)-\varPi_\eta\log_2(\gamma_\eta)),
%\end{align}

\begin{align}
\Bar{R}_\eta= \alpha_\eta\log_2(\gamma_\eta)+\beta_\eta,
\end{align}

where $\alpha_\eta=\hat{\gamma_\eta}/(1+\hat{\gamma_\eta})$ and $\beta_\eta=\log_2(1+\hat{\gamma_\eta})-\dfrac{\hat{\gamma_\eta}}{(1+\hat{\gamma_\eta})}\log_2(\hat{\gamma_\eta})$, where $\hat{\gamma_\eta}$ denotes the value of $\gamma_\eta$ from the previous iteration. Next, we compute NOMA power allocation at the LEO satellite, given the fixed phase shift design.

\subsection{NOMA Power Allocation}
For any given phase shift design at T-BD-RIS, the problem $\mathcal P_0$ can be simplified into a LEO NOMA power allocation problem such as follows:
%%%%%%%%%%%%%%%%%%%%%%%%%%%\\
\begin{alignat}{2}
\mathcal P_1:\begin{cases}  \underset{{p_i,p_j}}{\text{max}}\ (\Bar{R}_i+\Bar{R}_j)\label{7}\\
\text{subject to:} \\
C_1:\ \Bar{R}_{\eta}\geq R_{min},\ \eta\in i,j,\\
C_3:\ p_i+p_j\leq 1,\\
C_4:\ P_t(p_i+p_j)\leq P_{max},
\end{cases}
\end{alignat}
%%%%%%%%%%%%%%%%%%%%%%%%%%
The optimization problem $\mathcal P_1$ is the power allocation at LEO satellite and can be efficiently solved by KKT conditions. First, we define the Lagrangian function of $\mathcal P_1$ as:
\begin{align}
&L(p_i,p_j)=(\Bar{R}_i+\Bar{R}_j)-\lambda_i(\Bar{R}_{i}-R_{min})-\lambda_j(\Bar{R}_{j}-R_{min})\nonumber\\&-\mu_1(1-(p_i+p_j))+\mu_2(P_{max}-(p_i+p_j)P_t)
\end{align}
where $\lambda_i,\lambda_j,\mu_1,\mu_2$ denote the Lagrangian multipliers associated with the Lagrangian function. Next, we exploit KKT conditions to compute transmit power of $U_i$ first by calculating the partial derivative w.r.t $p_i$, which can be stated as:
\begin{align}
\frac{\partial L(\lambda_i,\lambda_j,\mu_1,\mu_2)}{\partial p_i}|_{p_i=p_i^*}=0,\label{12}
\end{align}
By expanding (\ref{12}), it can be described as:
\begin{align}
&=\frac{\partial}{\partial p_i}((\alpha_i\log_2(\gamma_i)+\beta_i)+(\alpha_j\log_2(\gamma_j)+\beta_j)\nonumber\\&-\lambda_i((\alpha_i\log_2(\gamma_i)+\beta_i)-R_{min})\label{13}\\&-\lambda_j((\alpha_j\log_2(\gamma_j)+\beta_j)-R_{min})\nonumber\\&-\mu_1(1-(p_i+p_j))-\mu_2(P_{max}-(p_i+p_j)P_t))=0,\nonumber
\end{align}
After calculating the partial derivation, the value of $p^*_i$ can be found as:
\begin{align}
    &p_i^*=\\&\dfrac{-\!\alpha_i \varPsi_j (\lambda_i\!-\!1) P_t \!+\!\alpha_j \varPsi_j (\lambda_j\!-\!1) P_t\!+\! (\mu_1 \!-\!\mu_2 P_t)\sigma^2 \!\!\pm\! \sqrt{\Lambda}}{2 \varPsi_j P_t (\mu_2 P_t \!-\!\mu_1)}, \nonumber
\end{align}
where

\begin{align}
    &\Lambda=4 \alpha_i \varPsi_j (\lambda_i-1) P_t (\mu_1-\mu_2 P_t)\sigma^2+(\alpha_i \varPsi_j (\lambda_i-1)\nonumber\\& P_t -\alpha_j \varPsi_j (\lambda_j-1) P_t-\mu_1 \sigma^2+\mu_2 P_t \sigma^2)^2\\
    &\varPsi_i=|{\bf h}_i\boldsymbol{\Phi}_t|^2\\
    &\varPsi_j=|{\bf h}_j\boldsymbol{\Phi}_t|^2
\end{align}
After getting $p^*_i$ for $U_i$, we can efficiently calculate the power of $U_j$ as:
\begin{align}
p^*_j=(1-p^*_i)P_t.
\end{align}

%%%%%%%%%%%%%%%%%%%%%%%%%%
\subsection{Phase Shift Design for T-BD-RIS}
Given the computed power values, $p^*_i$ and $p^*_j$ at the LEO satellite, the problem $\mathcal{P}_0$ can be transformed into a phase shift design optimization problem for the T-BD-RIS. The optimization problem can be formulated as follows:
\begin{alignat}{2}
\mathcal{P}_2:\quad &
\underset{\boldsymbol{\Phi}_t}{\text{max}} \ (\Bar{R}_i + \Bar{R}_j), \\ 
s.t. \quad &
C_2: \ \boldsymbol{\Phi}_t \boldsymbol{\Phi}_t^H = {\bf I}_K,
\end{alignat}
where $\boldsymbol{\Phi}_t \in \mathbb{C}^{K_x \times K_y}$ is the beyond diagonal phase shift matrix of the T-BD-RIS. The term $\Bar{R}_i + \Bar{R}_j$ represents the sum SE of the system, which depends on the phase shift design in the effective channel gains $|{\bf h}_i^H \boldsymbol{\Phi}_t|^2$ and $|{\bf h}_j^H \boldsymbol{\Phi}_t|^2$. The effective channel gain can be rewritten as:
\begin{align}
|{\bf h}_\eta^H \boldsymbol{\Phi}_t|^2 =  \text{Tr}(\boldsymbol{\Phi}^H_t {\bf h}_\eta {\bf h}_\eta^H \boldsymbol{\Phi}_t),\ \eta\in\{i,j\},\label{19}
\end{align}
Substituting this into the objective, the optimization problem becomes:
\begin{alignat}{2}
\mathcal{P}_{2.1}: \quad &
\underset{\boldsymbol{\Phi}_t}{\text{max}} \ (\hat{R}_i + \hat{R}_j), \\ 
s.t. \quad &
C_2: \ \boldsymbol{\Phi}_t \boldsymbol{\Phi}_t^H = {\bf I}_K.
\end{alignat}
where $\hat{R}_i=\alpha_i\log_2(\Gamma_i)+\beta_i$ and $\hat{R}_j=\alpha_j\log_2(\Gamma_j)+\beta_j$. The terms $\Gamma_i$ and $\Gamma_j$ can be described as:
\begin{align}
\Gamma_i = \frac{\text{Tr}(\boldsymbol{\Phi}^H_t {\bf h}_i {\bf h}_i^H \boldsymbol{\Phi}_t)p_iP_t}{\sigma^2} \\
\Gamma_j = \frac{\text{Tr}(\boldsymbol{\Phi}^H_t {\bf h}_j {\bf h}_j^H \boldsymbol{\Phi}_t)p_jP_t}{\sigma^2+\text{Tr}(\boldsymbol{\Phi}^H_t {\bf h}_j {\bf h}_j^H \boldsymbol{\Phi}_t)p_iP_t}
\end{align}
To solve $\mathcal{P}_{2.1}$ efficiently, we apply the semi-definite relaxation (SDR) method. Let us introduce a new matrix ${\bf W}$, where ${\bf W} = \boldsymbol{\Phi}_t \boldsymbol{\Phi}_t^H$. The optimization problem is relaxed by removing the rank-1 constraint on ${\bf W}$, yielding the following:
\begin{alignat}{2}
\mathcal{P}_{2.2}: \quad &
\underset{{\bf W}}{\text{max}} \ \alpha_i\log_2\left(\frac{\text{Tr}({\bf W}\textbf{F}_i)p_iP_t}{\sigma^2}\right)+\beta_i\\&+\alpha_j\log_2\left(\frac{\text{Tr}({\bf W}\textbf{F}_j)p_jP_t}{\sigma^2+\text{Tr}({\bf W}\textbf{F}_j)p_iP_t}\right)+\beta_j,\nonumber \\
s.t. \quad &
C_2: \ {\bf W} \succeq 0, \quad \text{Tr}({\bf W}) = K.
\end{alignat}
where $\textbf{F}_\eta={\bf h}_\eta {\bf h}_\eta^H$ with $\eta\in\{i,j\}$.
We can observe that the second term in the objective function is still non-convex. More specifically, we can write the second term as:
\begin{align}
\alpha_j \left[\log_2(\text{Tr}({\bf W}\textbf{F}_j)) - \log_2(\sigma^2 + \text{Tr}({\bf W}\textbf{F}_j)p_iP_t)\right] + \beta_j,
\end{align}
where $\log_2(\text{Tr}({\bf W}\textbf{F}_j))$ is concave in ${\bf W}$ and \(-\log_2(\sigma^2 + \text{Tr}({\bf W}\textbf{F}_j)p_iP_t)\) is convex in \({\bf W}\). Thus, the second term is a difference of concave and convex functions, which is not guaranteed to be convex or concave. To handle this, we can adopt first order Taylor approximation method on the convex part of the second term such as:
\begin{align}
g({\bf W}) = \log_2(\sigma^2 + \text{Tr}({\bf W}\textbf{F}_j)p_iP_t).
\end{align}
The first-order Taylor approximation of \(g({\bf W})\) around a feasible point \(\hat{{\bf W}}\) is:
\begin{align}
g({\bf W}) \approx g(\hat{{\bf W}}) + \nabla g(\hat{{\bf W}})^T ({\bf W} - \hat{{\bf W}}),
\end{align}
where
\begin{align}
g(\hat{{\bf W}}) = \log_2(\sigma^2 + \text{Tr}(\hat{{\bf W}}\textbf{F}_j)p_iP_t),
\end{align}
and the gradient \(\nabla g(\hat{{\bf W}})\) can be stated as:
\begin{align}
\nabla g(\hat{{\bf W}}) = \frac{p_iP_t}{\ln(2) (\sigma^2 + \text{Tr}(\hat{{\bf W}}\textbf{F}_j)p_iP_t)} \textbf{F}_j.
\end{align}
Using the Taylor approximation, the second term \(f({\bf W})\) can be written as:
\begin{align}
&f({\bf W}) \approx \log_2(\text{Tr}({\bf W}\textbf{F}_j)) \nonumber\\&- \left[g(\hat{{\bf W}}) + \nabla g(\hat{{\bf W}})^T ({\bf W} - \hat{{\bf W}})\right].
\end{align}
Now, substituting this into the original second term, we get:
\begin{align}
&\alpha_j \Big[\log_2(\text{Tr}({\bf W}\textbf{F}_j)) \nonumber\\& -  \left(g(\hat{{\bf W}}) + \nabla g(\hat{{\bf W}})^T ({\bf W} - \hat{{\bf W}})\right)\Big] + \beta_j.
\end{align}
After applying the Taylor approximation, the transformed problem in each iteration becomes:
\begin{alignat}{2}
\mathcal{P}_{2.3}: \quad & 
\underset{{\bf W}}{\text{max}} \ \alpha_i \log_2\left(\frac{\text{Tr}({\bf W}\textbf{F}_i)p_iP_t}{\sigma^2}\right) + \beta_i \\ & 
+ \alpha_j \Big[\log_2(\text{Tr}({\bf W}\textbf{F}_j))- \nonumber\\ &\left(g(\hat{{\bf W}}) + \nabla g(\hat{{\bf W}})^T ({\bf W} - \hat{{\bf W}})\right)\Big] + \beta_j, \nonumber\\
s.t. \quad &
C_2: \ {\bf W} \succeq 0, \quad \text{Tr}({\bf W}) = K.
\end{alignat}
The problem $\mathcal{P}_{2.3}$ can be solved using standard convex optimization tools (i.e., CVX), where ${\bf W}$ is approximated as a rank-1 matrix via eigenvalue decomposition as:
\begin{align}
{\bf W} = {\bf U} \boldsymbol{\Sigma} {\bf U}^H.
\end{align}
where ${\bf U}$ is a unitary matrix whose columns are the eigenvectors of ${\bf W}$. Each column of ${\bf U}$ corresponds to an eigenvector of ${\bf W}$, and these eigenvectors form an orthonormal basis $({\bf U}^H{\bf U}={\bf I})$. Note that in the context of the phase shift design, 
${\bf U}$ contains information about the directions in which the phase shifts of the BD-RIS are aligned. Moreover, $\boldsymbol{\Sigma}$ is a diagonal matrix whose diagonal elements are the eigenvalues of ${\bf W}$. These eigenvalues represent the magnitude of the signal energy in the directions defined by the eigenvectors in ${\bf U}$. The eigenvalues are non-negative because ${\bf W}$ is positive semi-definite. The optimal $\boldsymbol{\Phi}_t$ is then reconstructed as:
\begin{align}
\boldsymbol{\Phi}_t = {\bf U} \text{diag}(\sqrt{\boldsymbol{\Sigma}}).
\end{align}
To further refine the solution, the effective channel gains can be updated using the reconstructed $\boldsymbol{\Phi}_t$. The iterative process continues until convergence. By combining the SDR with iterative refinement, this approach efficiently designs the T-BD-RIS phase shift matrix to maximize the SE of the system while satisfying the QoS constraints for IoTDs.
%%%%%%%%%%%%%%%%
\begin{figure}[!t]
\centering
\includegraphics [width=0.5\textwidth]{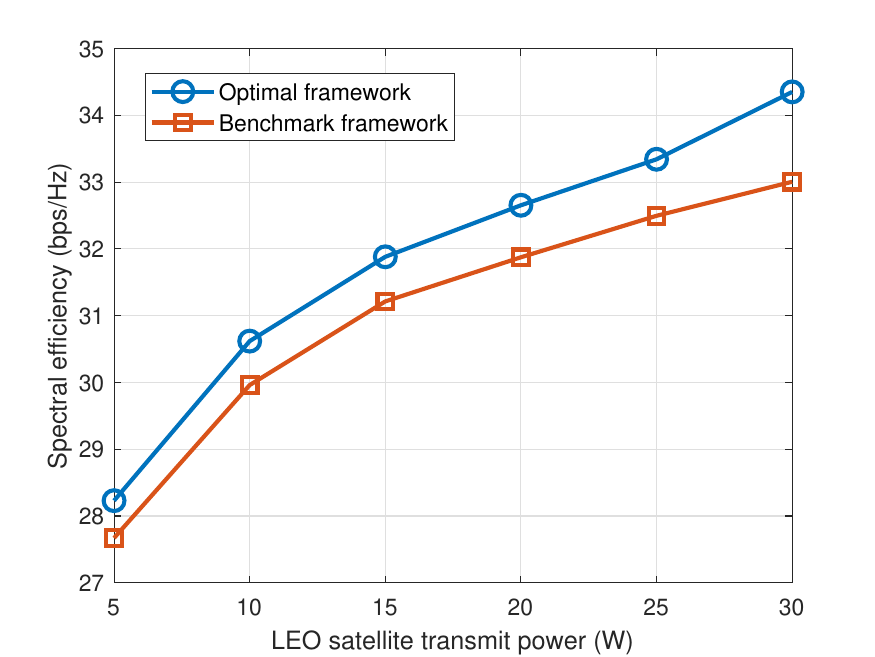}
\caption{Achievable spectral efficiency versus available transmit power of LEO satellite for T-BD-RIS for the proposed framework and the benchmark framework, where the phase shift elements of T-BD-RIS is set as 64.}
\label{Fig1}
\end{figure}
%%%%%%%%%%%%%

\section{Numerical Results}
This section provides the numerical results of the proposed optimization solution based on Monte Carlo simulations. We compare the proposed optimization (denoted as the optimal framework) with the benchmark solution (stated as the benchmark framework). In the benchmark framework, the spectral efficiency of the system is optimized with optimal T-BD-RIS phase shift design and fixed NOMA power allocation. Unless mentioned otherwise, the simulation parameters are set as follows: the transmit power of the LEO satellite is $P=20$ W, the phase shift elements of T-BD-RIS is $K=64$, the height of LEO is $500$ km, the carrier frequency is $f_c=18$ GHz (Ka-band), the bandwidth is 20 MHz, and the variance is $\sigma=0.00001$.

Fig. \ref{Fig1} illustrates the spectral efficiency versus the available LEO satellite's transmit power for the proposed optimal framework and the benchmark framework. The optimal framework, which employs both optimal NOMA power allocation and optimal T-BD-RIS phase shift design, consistently outperforms the benchmark framework, which relies on fixed NOMA power allocation with optimal T-BD-RIS phase shift design. As the transmit power increases from 5 W to 30 W, the spectral efficiency improves for both frameworks, showcasing the direct relationship between transmit power and system performance. Moreover, the performance gap between the two frameworks grows with the increasing transmit power of LEO, which shows the effectiveness of the proposed optimal framework. 

%%%%%%%%%%%%%%%%
\begin{figure}[!t]
\centering
\includegraphics [width=0.5\textwidth]{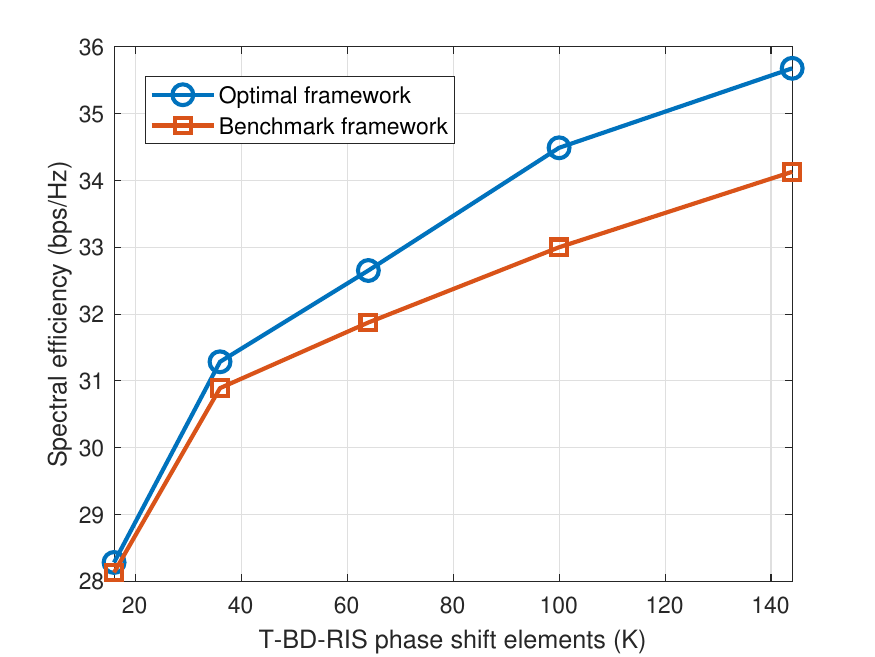}
\caption{Achievable spectral efficiency versus available phase shift elements of T-BD-RIS for the proposed framework and the benchmark framework, where the transmit power of LEO is set as 20 W.}
\label{Fig2}
\end{figure}
%%%%%%%%%%%%%
Fig. \ref{Fig2} demonstrates the spectral efficiency versus the number of T-BD-RIS phase shift elements for the proposed optimal framework and the benchmark framework, where the transmit power of LEO satellite is set as $P_t=20$ W. The optimal framework incorporates both optimal NOMA power allocation and optimal T-BD-RIS phase shift design, whereas the benchmark framework uses fixed NOMA power allocation with optimal T-BD-RIS phase shift design. As the number of phase shift elements increases from 16 to 144, the spectral efficiency improves for both frameworks, highlighting the crucial role of T-BD-RIS in enhancing the system's performance by improving the beamforming gains. The optimal framework consistently outperforms the benchmark framework across all values of $K$. The widening gap between the two frameworks as the phase shift elements increase emphasizes the effectiveness of the proposed joint optimization framework, which fully exploits the degrees of freedom provided by a larger number of T-BD-RIS elements. 
\section{Conclusion}
RIS technology has the potential to enhance the performance of satellite communication. This paper has proposed a new framework of T-BD-RIS mounted NOMA LEO satellite IoT networks. In particular, the proposed framework has optimized the power allocation of the LEO satellite and the phase shift design of T-BD-RIS to maximize the spectral efficiency of the system. The non-convex optimization was transformed using the SCA method and divided into two parts before obtaining an efficient solution. Then, a KKT-based closed-form solution for power allocation and SDR for phase shift design were exploited. Numerical results demonstrate the benefits of the proposed optimal framework compared to the benchmark framework. In the future, this work can be extended in several ways, i.e., this work can be extended to a multi-carrier NOMA communication; it can also be investigated by considering other performance metrics such as energy efficiency and secrecy performance. 

\bibliographystyle{IEEEtran}
\bibliography{Wali_EE}
\end{document}